# A Data Science Approach to Analyze the Association of Socioeconomic and Environmental Conditions With Disparities in Pediatric Surgery

Oguz Akbilgic [1]*, Eun Kyong Shin [2] and Arash Shaban-Nejad [3]*

[1] Department of Health Informatics and Data Science, Parkinson School of Health Sciences and Public Health, Loyola University Chicago, Chicago, IL, United States, [2] Department of Sociology, Korea University, Seoul, South Korea, [3] Department of Pediatrics, Center for Biomedical Informatics, College of Medicine, University of Tennessee Health Science Center, Memphis, TN, United States

**Background:** Scientific evidence confirm that significant racial disparities exist in healthcare, including surgery outcomes. However, the causal pathway underlying disparities at preoperative physical condition of children is not well-understood.

**Objectives:** This research aims to uncover the role of socioeconomic and environmental factors in racial disparities at the preoperative physical condition of children through multidimensional integration of several data sources at the patient and population level.

**Methods:** After the data integration process an unsupervised k-means algorithm on neighborhood quality metrics was developed to split 29 zip-codes from Memphis, TN into good and poor-quality neighborhoods.

**Results:** An unadjusted comparison of African Americans and white children showed that the prevalence of poor preoperative condition is significantly higher among African Americans compared to whites. No statistically significant difference in surgery outcome was present when adjusted by surgical severity and neighborhood quality.

**Conclusions:** The socioenvironmental factors affect the preoperative clinical condition of children and their surgical outcomes.

# INTRODUCTION

Surgical disparities became rapidly growing concerns both in health care and clinical services (1–4). Different social determinants of health (5) and experiencing unequal treatments (6) systematically produce, and further enhance, the disparities in health conditions and clinical outcomes. The high prevalence of poverty in the United States is a critical public health challenge. In 2018, approximately 38.1 million Americans lived in poverty (7). Memphis, Tennessee is consistently ranked among the poorest cities in the United States. Among racial and ethnic groups, African Americans had the highest poverty rate, 27.4%, followed by Hispanics at 26.6% and whites at 9.9%. Forty five point eight percent of young black children (under age six) live in poverty, compared to 14.5% of white children" (8).

Even with a dramatic decline in surgical mortality and morbidity of children in the United States over the past thirty years (9) still, African American children are at more than two-fold higher risk for death after surgery as white children (10). A recent study (11) revealed that African American children had 3.43 times the odds of dying within 30 days after surgery; 18% relative greater odds of developing postoperative complications, and 7% relative higher odds of developing serious adverse events. Furthermore, there are disparities at the preoperative condition of children that partially explains the disparities at surgery outcome (12). Yet, these observed racial disparities in surgical outcomes could be driven by underlying social disparities and the function of social conditions where an ethnic minority tends to be exposed to more vulnerable situations than others.

Sociomarkers, "measurable indicators of social conditions in which a patient is embedded" (13) can play a critical role in predicting many health outcomes including adverse surgical outcomes. Memphis, Tennessee with roughly 65% of the African American population experiences one of the highest rates of health disparities across the nation. African Americans experience worse health status on multiple indicators in comparison with whites (14) mainly because of disparities in socioeconomic and environmental conditions. Therefore, socioenvironmental constructions have major implications for how measures of race contribute to human health (15). Not only differences at the individual level but also disparities at the neighborhood level sociomarkers can be factored into the reproduction mechanism of health inequalities (13).

Despite the importance of sociomarkers, there are few data available to date examining the direct associations between the sociomarkers and clinical risk factors linked to surgery outcomes, especially in the pediatric population. The causality links between population-level sociomarkers and patient-level clinical risk factors are, thus, not well-understood. By integrating surgical outcome data with socioenvironmental data, the present work unpacks the underlying mechanisms of the radical disparities in risk for death after surgery among pediatric populations. This study aims to investigate the role of socioenvironmental factors of racial disparities at the preoperative physical condition of children through multidimensional integration of several relevant data sources.

## METHODS

We hypothesized that racial disparities in postoperative outcomes can be explained with disparitis in preoperative and the living conditions of children undergoing surgery. To test our hypothesis, we have integrated clinical data from pediatric surgery cases with population-level neighborhood quality metrics from government and private sector resources.

## Cohort

Le Bonheur Children's Hospital (LBCH), Memphis, TN participates in the National Surgical Quality Improvement Project-Pediatric (NSQIP-Ped) (16), and a trained surgical case reviewer abstracts clinical data for a non-random sample of procedures corresponding to about 15% of all surgeries. In this study, we analyzed children undergoing surgical procedures at on or before their 19th birthday, whose operation occurred between January 1, 2012, and December 31, 2017, their data were abstracted for NSQIP-Ped.

## Population-Level Data

We have used zip code data associated with the surgical cases in our cohort to merge clinical data with population-level data. We utilized zip code level 2010 US Census data and 244,000 property and neighborhood quality data collected by the Memphis Property Hub (mempropertyhub.com), a private sector partner that collects data related to distressed and vacant properties in the 29 zip codes in the City of Memphis, TN.

## Clustering

We implemented a k-means clustering algorithm (17) to group 29 zip codes into two categories by setting k = 2. k-means is an unsupervised clustering algorithm aiming to split observations into k-groups by finding k centers in the feature space where each observation is associated with the closest center. The center points are iteratively changed to find their optimal location where within-cluster variance is minimized while maximizing variance across centers. Therefore, in our case, the k-means algorithm will automatically group zip codes with similar socioeconomic characteristics in the same cluster. One of the main advantages of the k-means algorithm is that it is an unsupervised method not exposed to overfitting obtained due to supervised classification algorithms.

## Disparities at the Preoperative Physical Condition of Children

We previously developed a pediatric surgery risk stratification model (10, 18) utilizing preoperative clinical risk factors at the time of surgery. Our risk strata were built, trained, and validated using over 250,000 individual pediatric surgical cases from NSQIP-Ped 2010–2015 dataset. In this study, we used our risk strata to represent the preoperative physical condition of children. Our risk stratification model assigns surgical cases into one of five risk categories ranging from A to E with increased risk for death. The percentage risk of death with 95% confidence intervals for these five risk strata are 0.08 (0.06, 0.09), 1.35 (0.92, 1.77), 3.30 (2.72, 3.88), 12.55 (10.85, 14.26), 38.60 (33.74, 43.46), respectively. Using preoperative risk factors, we identify the corresponding risk strata for all patients in our cohort and further combined risk groups B, C, D, and E in the same group representing Poor Preoperative Condition. Therefore, risk strata A [children who had no ventilator dependency and did not receive oxygen support 48 h prior to surgery and who did not have do-not-resuscitate (DNR) order] represents the children and Good Preoperative Condition.

**Transfer status** is a variable summarizing the setup that the patients are coming to the operating room from. Transfer Status 1 is for patients coming to the operating room from home, clinic, or doctor's office. Transfer Status 2 is for patients coming to the operating room from the emergency room including outside emergency rooms with direct hospital readmission. Transfer 3 is for patients coming to the operating room from an outside hospital, Neonatal intensive care unit (NICU), Pediatric intensive care unit (PICU), inpatient general floor, or adult ICU. Among all 2,638 cases, 1,512 cases had Transfer Status 1, 925 Transfer Status 2, and 201 Transfer Status 3. Despite there may be exceptions, in general, from Transfer Status 1 to 3, the prevalence of adverse surgery outcomes increases. For instance, 0.20% of Transfer 1, 1.73% of Transfer Status 2, and 35.32% if Transfer Status 3 patients had extended hospital stay beyond 30 days.

TABLE 1 | Clustering zip codes into poor and high-quality neighborhoods.

| Socioeconomic and environmental factor | Zip code cluster | Mean (95% CI) |
| --- | --- | --- |
| Unoccupancy (%) | 1 (High Quality) | 2.48 (1.48, 3.47) |
|  | 2 (Poor Quality) | 10.89 (7.93, 13.84) |
| Broken windows (%) | 1 (High Quality) | 0.12 (0.00, 0.24) |
|  | 2 (Poor Quality) | 1.18 (0.84, 1.51) |
| Dumping (%) | 1 (High Quality) | 0.05 (0.02, 0.07) |
|  | 2 (Poor Quality) | 0.60 (0.32, 0.88) |
| Education attainment (%) | 1 (High Quality) | 93.21 (91.46, 94.96) |
|  | 2 (Poor Quality) | 77.22 (74.22, 80.21) |
| Individual below poverty (%) | 1 (High Quality) | 10.04 (6.90, 13.18) |
|  | 2 (Poor Quality) | 37.19 (32.84, 41.53) |
| Neighborhood Inequality | 1 (High Quality) | 0.54 (0.47, 0.61) |
|  | 2 (Poor Quality) | 0.87 (0.79, 0.96) |

## Statistical Tests

We implement several statistical hypothesis tests to determine whether there are racial disparities in the preoperative condition of children or not. We used a Chi-Square test to test for homogeneity of 2 × 2 contingency and Fisher's Exact Test when any of the frequency within the contingency tables were smaller than five. We implemented the Mann-Whitney U test to compare continuous variables coming from two categories and the independent sample Z-test to compare two proportions.

## RESULTS

### Cohort

Our cohort is of 2,638 surgical cases including 58.8% male with average age and standard deviation (SD) of 7.2 ± 5.9 and with the racial distribution of 55.4% African American, 26.5% White, 16.8% unknown/unreported, and 1.3% others. The average age and SD for African Americans and white are 7.7 ± 5.9 and 7.2 ± 6.1.

### Neighborhood Characteristics

We implemented an unsupervised k-means clustering algorithm to group zip codes into two categories by setting k = 2. Fourteen zip codes clustered in Cluster 1, corresponding to 1,082 cases, and 15 zip codes in Cluster 2, corresponding to 1,556 cases, with their characteristics summarized in **Table 1** and **Figure 1**.

**Table 1** and **Figure 1** clearly show that neighborhood quality metrics are significantly better (both Interdepend Sample T and Mann Whitney $U$ tests $p$-value < 0.001) for Cluster 1 zip codes (High-Quality Neighborhood) compared to Cluster 2 zip codes (Poor Quality Neighborhood). The proportion of the number of African American to the number of White population within the good quality neighborhood is 1.23 (95% CI, 0.07, 2.38) while it is significantly ($p$ < 0.01) higher in poor quality neighborhoods, 12.82 (95% CI, 5.00, 20.63).

We compared the two neighborhood clusters using Chi-Square Test in terms of adverse surgical outcomes such as 90-day postoperative mortality, unplanned readmission, extended hospital stays beyond 30 days, and postoperative complications. We found that all the outcomes in consideration were significantly more prevalent (unadjusted Chi-Square test $p$ = value of <0.05) in surgical cases operate on children from poor-quality neighborhoods.

### Disparities at the Preoperative Physical Condition of Children

We compared the Preoperative Condition (Good or Poor) of African American and white children using Chi-Square test and found that African American children are in significantly worse physical condition at the time of surgery compared to white children ($p$ < 0.001). Six point five seven percent of African American children and 2.71% of white children were at Poor Physical Condition at the time of surgery. However, we note that these are unadjusted comparisons. Therefore, we implemented two more levels of adjustments in the cohort. First, we adjusted patients by a variable representing surgical severity, and then, we adjusted patients with the neighborhood quality. In the first step, we used the transfer status variable and in the second step, we used the two neighborhood clusters (poor or good).

As a first step, we compared African American and white children in terms of their Preoperative Physical Condition adjusted by transfer status in **Table 2**. Our results showed that African American children were at the worse preoperative physical condition at the time of surgery for only Transfer Status 2 cases (emergency room transfers).

In the second step, we further adjusted the cohort by the neighborhood quality in **Table 3** for only transfer Status 2 cases. **Table 3** shows that there is no significant difference between African American and white children living in good quality neigborhood ($p$ = 0.502) once it is adjusted by Transfer Status and Neighborhood Quality. It was the case for poor quality neigborhoods with smaller yet still not signicant $p$-value of 0.91. Taking social and neighborhood conditions into accounts, we do not observe any significant racial disparities in this group of pediatric patients.

## DISCUSSION

African American children disproportionately live in poor-quality neighborhoods. The quality of the neighborhood and the residential environment is associated with the level of happiness, diet and physical activities, and many health outcomes (19, 20). Poverty, which entails lack of resources, prevents individuals from living with dignity and jeopardizes their health and wellness. Disparities in surgical outcomes are not exempt from this chronic public health problem.

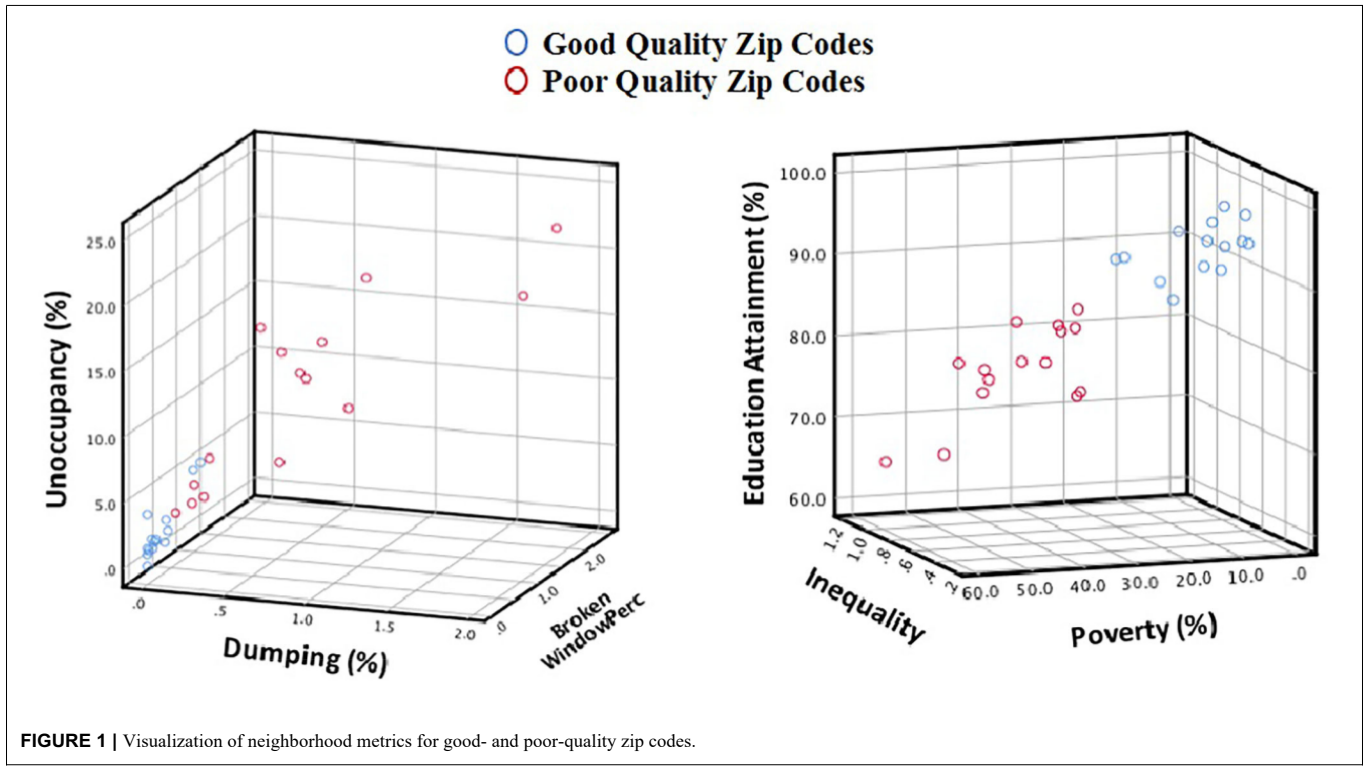

FIGURE 1 | Visualization of neighborhood metrics for good- and poor-quality zip codes.

TABLE 2 | Comparison of African American and White children by their preoperative physical condition adjusted by transfer status.

| Transfer Status | Race | Preoperative physical condition | | p-value (Chi-Square or fisher's exact test) |
|---|---|---|---|---|
| | | Good | Poor | |
| 1 | White | 426 | 2 | 0.240 |
| | African American | 854 | 11 | |
| 2 | White | 234 | 5 | 0.034 |
| | African American | 437 | 25 | |
| 3 | White | 21 | 12 | 0.400 |
| | African American | 75 | 60 | |

TABLE 3 | Comparison of African American and White children by their preoperative physical condition adjusted by transfer status and neighborhood quality.

| Neighborhood quality | Transfer Status | Race | Preoperative physical condition | | P-value (Chi-Square or fisher's exact test) |
|---|---|---|---|---|---|
| | | | Good | Poor | |
| Good | 2 | White | 182 | 5 | 0.502 |
| | | African American | 95 | 4 | |
| Poor | 2 | White | 52 | 0 | 0.091 |
| | | African American | 342 | 21 | |

Health, equity, and justice are key elements in every prosperous human society (21). To improve the quality of healthcare and reduce disparities a clear understanding of the disease pathway (22), from upstream risk factors and health determinates to downstream health consequences and outcomes is needed. For that, the public health sector needs an innovative approach to integrate multi-dimensional data from different domains (e.g., social science, epidemiology, medicine, etc.) at different levels (individual, administrative, and population levels).

Data science has been widely used in understanding and addressing health disparities (23–26). In their study, Want et al. used machine learning algorithms to predict cardiovascular-related mortality risk and revealed that socio-behavioral factors were associated with increased risk. Singh and Yu (27) implemented log-linear regression to understand the role of racial, ethnic, and socioeconomic disparities in infant mortality trends. There are also data science approaches to investigate the role of health disparities in the surgery domain. Hanson et al. (28) used the random forest algorithm to study the importance of race and social factors in mortality following prostate cancer. In Akbilgic et al., authors used decision trees and network analysis to understand preoperative patterns of disparities in postoperative complications (10) among children. Despite this study revealed that disparities in postoperative complications were explained by disparities in preoperative physical fitness of children, it did not explain the reasons for disparities at the preoperative level. In this study, we aimed to use a data science approach to fill this gap and understand the underlying socioeconomic factors of disparities in the preoperative level in pediatric surgeries.

Adjusting patients with their neighborhood and insurance type can in part explain the racial disparities gap in the preoperative physical condition of children in Memphis, TN. Inference of our findings is constrained with some limitations and the results must be interpreted based on the scope condition. We considered the granularity of the zip code, which might not be the finest granular to capture detailed social gradients (29), for our neighborhood investigation. Although in our study, the zip code works relatively well, given the low population density in the subject area. To improve our causal inference investigating neighborhood effects on children's health we still need to bring much more granular data sets into our analysis, as well as build up theories and methodologies.

Our study has some limitations. First, our results represent data from a single hospital in Memphis, TN. Further, socioeconomic risk factors are used at the zip code level is limiting our capacity to capture heterogeneous socioeconomic structures within the same zip codes. Therefore, there is a need for a future study at more granular geographic data such as census tract level and validation on different cohorts.

## CONCLUSION

Our results show that a large proportion of racial disparities at the preoperative physical condition of children can be explained by socioeconomic and environmental factors. This is a call to action to take the next steps to study long-term outcomes and impacts in these children and generate actionable insights for informed decision making and policy development. However, there is a need for expanding this research by obtaining additional socioeconomic and environmental data at a more granular level such as the census tract or even at the individual level.

## DATA AVAILABILITY STATEMENT

The data analyzed in this study is subject to the following licenses/restrictions: The dataset would be available upon request and subject to the instutational IRB approval. Requests to access these datasets should be directed to oakbilgic@luc.edu.

## AUTHOR CONTRIBUTIONS

OA conceptualized and designed the study, collected, reviewed, curated data, and drafted the manuscript. ES conducted the analysis, curated data, and contributed in writing the manuscript for the introduction, methods and results. OA and AS-N conceived the study, received funding, participated in the study design, and revised the manuscript. All authors read and approved the final manuscript.

## FUNDING

This study was supported through a grant from the Children's Foundation Research Institute at Le Bonheur Children's Hospital.

## ACKNOWLEDGMENTS

We would like to thank Dr. Max Langham, Jr and Dr. Robert L. Davis from the University of Tennessee Health Science Center, Memphis, TN for their valuable insights, comments, and recommendations.